\documentclass[pra,showpacs,raggedbottom,twocolumn,superscriptaddress,floatfix,amsmath,A4]{revtex4}
\usepackage[latin1]{inputenc}
\usepackage[english]{babel}
\usepackage[T1]{fontenc}

\usepackage{graphicx}
\usepackage{amssymb}
\usepackage{amsmath}
\usepackage{amscd}
\usepackage{eucal}
\usepackage{color}
\usepackage{bm}

\usepackage[normalem]{ulem}

\newcommand{\e}{{\rm e}}
\newcommand{\rmd}{{\rm d}}
\newcommand{\rmi}{{\rm i}}

\newcommand{\half}{{\textstyle{\frac{1}{2}}}}

\newcommand{\quarter}{{\textstyle{\frac{1}{4}}}}

\newcommand{\de}{\delta}

\newcommand{\om}{\omega}

\newcommand{\He}{$^3$He}

\newcommand{\reffirstexactresult}{22}
\newcommand{\reflastexactresult}{27}

\newcommand{\refEqtelegrapher}{1}
\newcommand{\refEqresultaverage}{2}

\newcommand{\refEqresultcovar}{4}

\newcommand{\refFigdistribution}{2}

\newcommand{\citeFellerprobbook}{[5]}
\newcommand{\citeJacob}{[48]}
\newcommand{\citeJacobSchmiedeskamppartthree}{[48,50]}
\newcommand{\citeSchmiedeskamppartthree}{[50]}
\newcommand{\citeSchmiedeskamppartonetwo}{[49]}

\definecolor{DarkGreen}{rgb}{0,0.7,0}

\begin{document}

\title{ 
Size-independence of statistics for boundary collisions of random walks\\
and its implications for spin-polarized gases
}

\author{Dominique J.~Bicout}
\affiliation{
Institut Laue-Langevin,
6 rue Jules Horowitz, B.P.~156, 38042 Grenoble, France.}
\affiliation{
Biomath\'ematiques et Epid\'emiologie, EPSP-TIMC, UMR
5525 CNRS, Universit\'e Joseph Fourier and  VetAgro Sup, Campus
V\'et\'erinaire de Lyon, 1 avenue Bourgelat, 69280 Marcy l'Etoile,
France}

\author{Efim I.~Kats}
\affiliation{
Institut Laue-Langevin,
6 rue Jules Horowitz, B.P.~156, 38042 Grenoble, France.}
\affiliation{
Landau Institute for Theoretical Physics, Moscow, Russia.}

\author{Alexander K.~Petukhov}
\affiliation{
Institut Laue-Langevin,
6 rue Jules Horowitz, B.P.~156, 38042 Grenoble, France.}

\author{Robert S.~Whitney}
\affiliation{
Laboratoire de Physique et Mod\'elisation des Milieux Condens\'es (UMR 5493), 
Universit\'e Grenoble 1, Maison des Magist\`eres, B.P.~166, 38042 Grenoble, France.}
\affiliation{
Institut Laue-Langevin,
6 rue Jules Horowitz, B.P.~156, 38042 Grenoble, France.}

\date{\today}
\begin{abstract}
A bounded random walk exhibits strong correlations between collisions with a boundary.
For an one-dimensional walk, we obtain the full statistical distribution of the number of such collisions 
in a time $t$.
In the large $t$ limit, the fluctuations in the number of  collisions are found to be
{\it size-independent}  (independent of the distance between boundaries).
This occurs for any inter-boundary distance, including less and greater than the 
mean-free-path, and means
that this boundary effect does not decay with increasing system-size.
As an application, we consider spin-polarized gases, such as 3-Helium, in the three-dimensional 
diffusive regime.
The above results mean that the depolarizing effect of rare magnetic-impurities
in the container walls is orders of magnitude larger than a Smoluchowski assumption 
(to neglect correlations) would imply.  This could explain   
why depolarization is so sensitive to the container's treatment with magnetic fields prior to its use.

\end{abstract}
\maketitle

\begin{figure}
\includegraphics[width=\columnwidth]{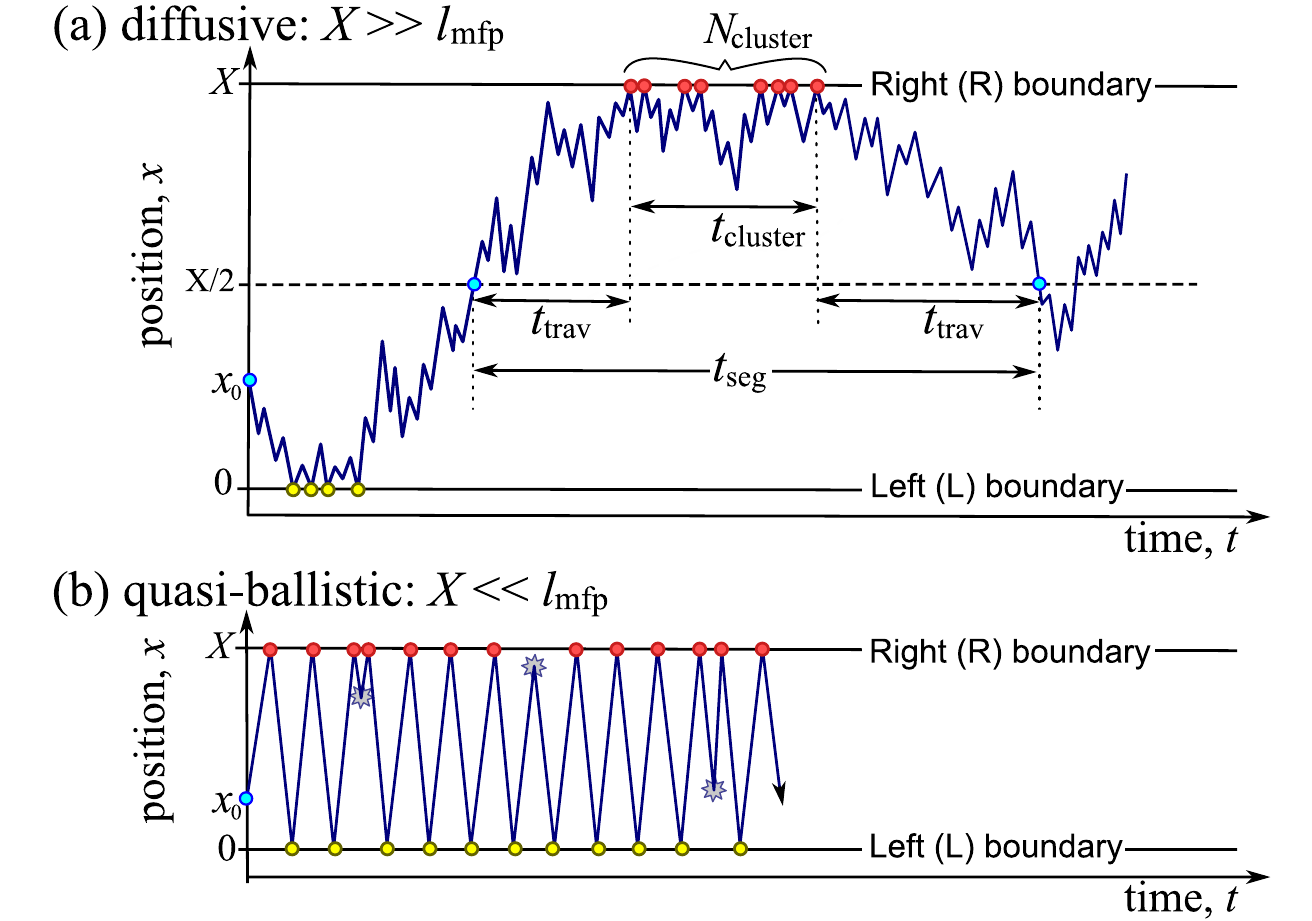}
\caption{\label{Fig:intuitive} 
Sketches of bounded random-walks:
(a) a diffusive walk and (b) a quasi-ballistic walk.  
In (a) the boundary-collisions are clustered. 
In (b) the boundary collisions are anti-clustered (predominantly equally spaced), because
random changes of direction (highlighted with stars) are rare.
}
\end{figure}

{\bf Introduction.}
Random-walks between (or near) boundaries crop-up throughout the mathematical sciences,
from diffusion of particles in a box,  
to biological systems
\cite{Jeanson2003,Pigolotti2005,Chapman2007,Olson2008}, 
or the gambler's ruin problem \cite{Feller-prob-book,Redner-book}.  
The last of these is a first passage problem (the probability that the walk hits the boundary at its $n$th step).
For a standard one-dimensional (1D) random walk, this first passage problem
has long since been solved \cite{Feller-prob-book,Redner-book}, 
and now most work is for higher dimensions in various geometries
\cite{Condamin2005-07,Deaconu2006,Mattos2012}, 
 or anomalous walks 
 \cite{Metzler2000,Zoia2007,Condamin2008,Burch2011}. 
 Other works study dynamics of walks within finite or bounded regions 
 \cite{Bicout1998,El-Shehaway2000,Zoia2011}.
Here we address a closely related problem; 
we take a standard 1D random walk trapped between two boundaries (labelled ``L'' and ``R''), 
and study the statistical distribution of the number of boundary-collisions 
in a time $t$. 
Despite a well known formal connection to the {\it recurrence time} (first-passage time
for a walk starting at the boundary) \cite{Feller1949,Feller-prob-book},
this problem does not appear to have been solved before now. 
We obtain the full distribution for arbitrary $t$,
and study the large $t$ limit.

Our central results, Eqs.~(\ref{Eq:result-var},\ref{Eq:result-covar}), show that the statistical fluctuations in the number of L and R  boundary-collisions,
$N_{\rm L}$ and $N_{\rm R}$,
exhibit {\it size-independence} for large time $t$.
Remarkably, this is a boundary effect which does not decay with increasing system size, $X$,
even though the average number of such collisions decays like $1/X$. 
The $X$-independences applies for all $X$, both greater
and less 
than the walk's mean-free-path, $l_{\rm mfp}$; 
see Fig.~\ref{Fig:intuitive}a and b.
It is a consequence of the correlations between subsequent boundary collisions,
with collisions clustering for large $X$ and anti-clustering for small $X$.
The clustering for large $X$ means the variance of $N_{\rm L,R}$  is very much greater than the average.
The only requirement is that $t$ is 
much larger than the mean-free-time 
$\tau_{\rm mfp}=l_{\rm mpf}/v$ (so the motion is random)
and the time to traverse the system, $t_{\rm trav}$ (so the walk explores the whole system).

Below we discuss
spin-polarized gases ({\He}, Xe, etc), which are  used in a variety
of scientific and medical situations, and argue that the above size-independent fluctuations
are a crucial source of depolarization.


{\bf Model and Results.}
A telegrapher equation is a standard model for one-dimensional random walkers
with a finite mean free path, $l_{\rm mfp}$ \cite{Feller-prob-book,Weiss2002,Kolesnik-Pinsky2011}. 
It models a particle moving with velocity $v$ in one-dimension, 
which changes the direction of its motion at random on a timescale $\tau_{\rm mfp}=l_{\rm mfp}/v$.   
We assume the walker is between two boundaries (see Fig.~\ref{Fig:intuitive}),
the left (L) boundary at $x=0$ and the right (R) boundary at $x=X$.
The walk reflects whenever it hits the boundary, i.e. its velocity is reversed. 
Working with dimensionless variables, $\tau =t/\tau_{\rm mfp}$ and $z= x/X$, 
we define $P_\pm(z,\tau)$ as the probability densities that the random-walker is at $z$ and 
in state $+$ or $-$ (i.e.~has velocity $+v$ or $-v$) at time $\tau$.
The telegrapher equation is equivalent to  \cite{Feller-prob-book,Weiss2002}, 
\begin{eqnarray}
{\textstyle {\rmd \over \rmd \tau}}P_\pm(z,\tau) = \mp\varepsilon {\textstyle{\rmd \over \rmd z}} P_\pm(z,\tau) - P_\pm(z,\tau)+P_\mp(z,\tau)\:, \ \
\label{Eq:telegrapher}
\end{eqnarray}
where $\varepsilon=v \tau_{\rm mfp}/X$. 
Eq.~(\ref{Eq:telegrapher}) is a pair of master equations
for the left or right motion of the walker.  
The first term on the right-hand-side is motion in the direction of travel
(with velocity $\varepsilon$ in our dimensionless units),
while the last two terms account for the changes of direction which occur at random
(at an average rate equal to one in our dimensionless units). 
The diffusive and quasi-ballistic limits 
are $\varepsilon\ll 1$ and $\varepsilon\gg 1$,
respectively.

For a walk starting at $z_0$ at $\tau=0$, we have
$P_\pm (z,0)= \half [1\mp (1- 2a)] \delta(z-z_0)$,
where $a$ is the probability that the initial velocity is positive.
We write the boundary conditions at  $z=0,1$ as
$P_+ (0, \tau)=\mu_{\rm L} P_-  (0, \tau)$ and  
$P_- (1,\tau)=\mu_{\rm R} P_+  (1, \tau)$,
where  $\mu_{\rm L}$ and $\mu_{\rm R}$ give escape probabilities at each boundary collision.
In reality $\mu_{\rm L}=\mu_{\rm R}=1$, however we leave them free so we can use them as counting variables 
to track the boundary collisions
\cite{Rubin1982,Bicout1999}.  
The survival probability \cite{Feller-prob-book} (probability to be in the system at time $\tau$), is 
$\Psi(\tau|z_0) =  \int_{0}^{1}\!dz\big[ P_-(z,\tau|z_0) +P_+(z,\tau|z_0)\big] $.

The Supplementary Material details the calculation~\cite{Feller-prob-book}
of the Laplace transform of the boundary-collision statistics,
with $\widehat{f}(s) = \int_0^\infty \rmd t \, \e^{-s\tau} \,f(\tau)$ for any $f(\tau)$.
To get the long time behavior, we do as follows.
\begin{enumerate}
\item
We show that $\widehat{\Psi}(s|z_0)$ is a generating function for the desired quantities
(generated by taking derivatives of $\widehat{\Psi}$ with respect to $\mu_{\rm L,R}$),
and that $\widehat{\Psi}(s|z_0)$ is given by $\widehat{P}_\pm(z,s|z_0)$ at $z=0,1$.

\item
We get $\widehat{P}_\pm (z,s|z_0)$ at $z=0,1$ by
writing Eq.~(\ref{Eq:telegrapher})'s Laplace transform as a matrix equation, 
using a Fourier transform in $z$ and a matrix diagonalisation.
We thereby get the exact algebraic expressions for 
$\big\langle \widehat{N}_{\rm L,R} (s|z_0)\big\rangle$, $\big\langle \widehat{N}_{\rm L,R}^2 (s|z_0)\big\rangle$, etc, given
in Eq.~(\reffirstexactresult-\reflastexactresult) of the Supplementary Mxaterial.

\item
For small $s$, we invert the Laplace transforms, to
get the statistics for $t \gg {\rm max}[\tau_{\rm mfp}, t_{\rm trav}]$.
\end{enumerate}
We also find that $t_{\rm trav}\simeq X/v + X^2/ (v^2 \tau_{\rm mfp})$. 
Thus, for $t \gg {\rm max}[\tau_{\rm mfp}, t_{\rm trav}]$, the above method gives us
\begin{eqnarray}
\big\langle N_{\rm L}\big\rangle = 
\big\langle N_{\rm R}\big\rangle &=& {vt/(2X)},  
\label{Eq:result-average}
\\
{\rm var}[N_{\rm L}] = 
{\rm var}[N_{\rm R}] &=&  t / (3\tau_{\rm mfp}), 
\label{Eq:result-var}
\\
{\rm covar}[N_{\rm L},N_{\rm R}]  &=&   -t / (6\tau_{\rm mfp}). 
\label{Eq:result-covar}
\end{eqnarray}
Since the distribution of $N_{\rm L,R}$ is gaussian \cite{Feller-prob-book}, 
all higher moments are given by the above second-moments, and so exhibit the same 
$X$-independence.
Eqs.~(\ref{Eq:result-average}-\ref{Eq:result-covar}) apply 
for any $X$, from quasi-ballistic ($X \ll l_{\rm mfp}$) to diffusive ($X \gg l_{\rm mfp}$).
Although for large $X/l_{\rm mfp}$ the time at which the dynamics enters the above long time limit
goes like $t_{\rm trav}\propto X^2$. 
In this diffusive limit, the variance is vastly larger than the average,
although typical fluctuations are of order $\sqrt{{\rm var}}$, and remain much less than
the average. 

The  index of dispersion is the variance-to-mean ratio (VMR), and it tells us about clustering
\cite{Clustering}.  For the L or R boundary collisions, the index is
\begin{eqnarray}
{{\rm var}[N_{\rm L,R}] \over \big\langle N_{\rm L,R}\big\rangle } = {2X \over 3l_{\rm mfp}}
\ \left\{ \!\!\begin{array}{ll} 
\gg 1 \hbox{ (clustered)}     &  \! \! \hbox{for } X\gg l_{\rm mfp} \\
= 1 \hbox{ (Poisson-like)}   &  \! \!  \hbox{for } X = {3 \over 2} l_{\rm mfp} \\
\ll 1 \hbox{ (anti-clust.)} &  \! \! \hbox{for } X\ll l_{\rm mfp} \\
\end{array}\right. \!\!\!,  \ 
\label{Eq:VMR}
\end{eqnarray}
 Fig.~\ref{Fig:intuitive}a,b show the clustering for $X\gg l_{\rm mfp}$ and anti-clustering for $X\ll l_{\rm mfp}$. 
Remarkably, Eq.~(\ref{Eq:result-var}) is exactly the same in two such physically different limits.

\begin{figure}
\includegraphics[width=\columnwidth]{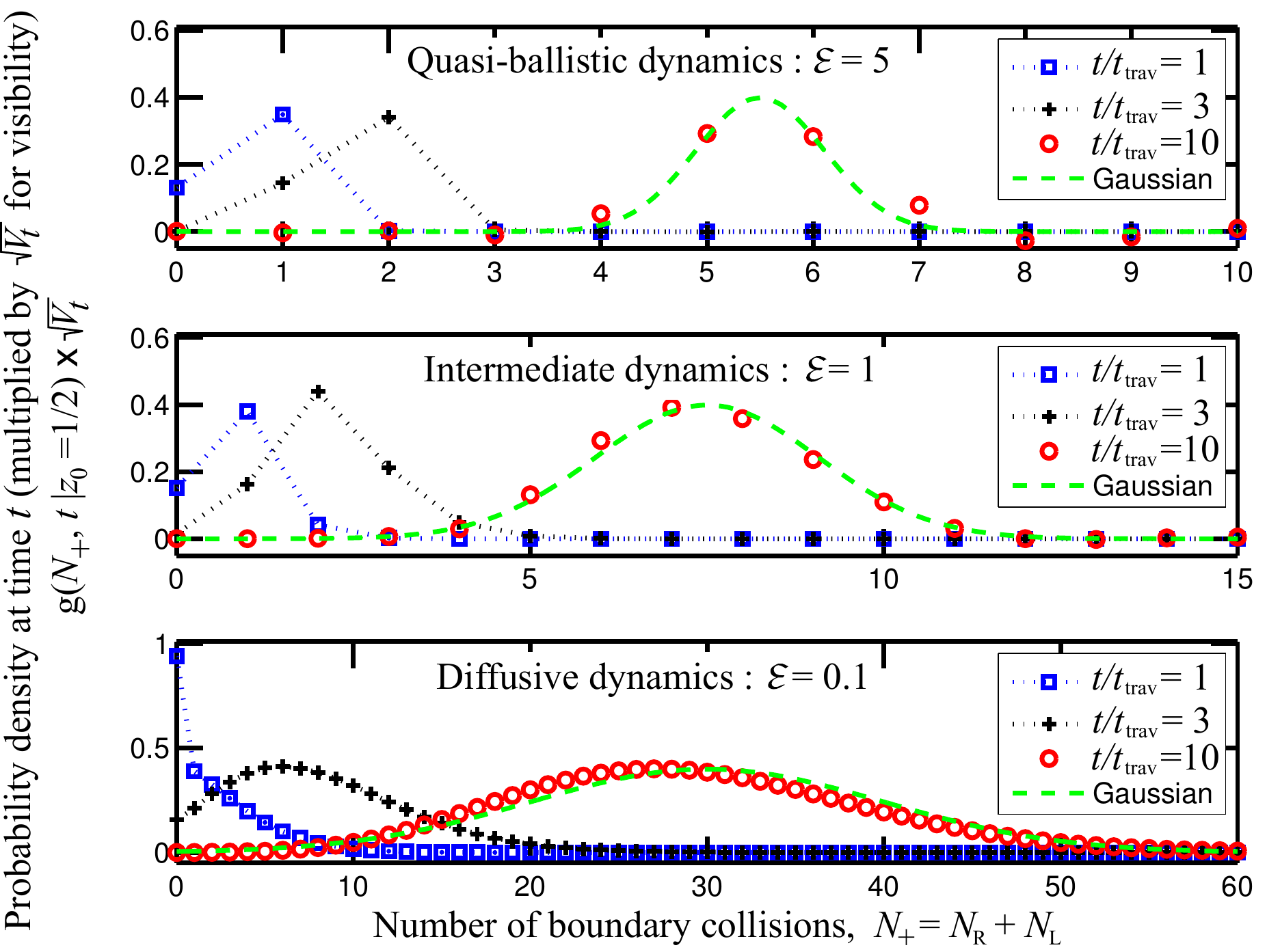}
\caption{\label{Fig:distribution}
The evolution of the distribution of boundary collisions with time, for three different $\varepsilon$,
given by numerically inverting the Laplace transformed distribution
in Eqs.~(\ref{Eq:distribution-first}-\ref{Eq:distribution-last}).
We plot the distribution of $N_+$,
for a particle initially at the mid-point between the boundaries,  $z_0=a=1/2$.
To make the time-evolution of the {\it shape} of the distribution clearly visible, 
we multiply the vertical scale by the width of the distribution, $V_t^{1/2}$, 
at each time $t$, where $V_t={\rm var}[N_+(t)]$.
The gaussian is that given below Eq.~(\ref{Eq:distribution-last}) for $t/t_{\rm trav}=10$.    
}
\end{figure}

We also get the statistical distribution of 
boundary-collisions for arbitrary times,
as an algebraic expression for its Laplace transform.  The Laplace transform of the probability density that a walk initially at $z_0$ experiences $N_+=N_{\rm L} + N_{\rm R}$
boundary collisions in a time $\tau$, is 
$\widehat{g}(N_+,s|z_0)
=(N_+!)^{-1}(\rmd/\rmd \mu)^{N_+} \widehat{\Psi}(s|z_0)\big|_{\mu=0}$, 
where we take $\mu_{\rm L}=\mu_{\rm R}=\mu$.
Since $\widehat{\Psi}$ is a fairly simple function of $\mu$, one can evaluate the derivatives
for any $N_+$.  Then
\begin{eqnarray}\label{gofn}
    \widehat{g}(N_+\! >\!0,s|z_0)
     = \! \sum_{\nu=\pm} \!
    \frac{\big[m(a)+u_\nu \,m(1\!-\!a) \big] (1-u_\nu)}{2\lambda\varepsilon
    u_\nu^{N_+ +1}s}, 
\label{Eq:distribution-first}
\end{eqnarray}
and $\widehat{g}(N_+\! =\!0,s|z_0) =s^{-1}(1-2m(a)/D_0)$, 
where 
\begin{eqnarray}
u_{\pm} &=& { {\sinh\left[\lambda\right]\pm\lambda\varepsilon} \over
(s+1)\sinh\left[\lambda\right]-\lambda\varepsilon\,\cosh\left[\lambda\right] },
\\
m(\alpha) 
&=& (1+(1-\alpha)s)\sinh\left[\lambda(1-z_0)\right]
\nonumber \\
& &
\ \ +(1+\alpha s)\sinh\left[\lambda z_0\right] 
+ \alpha\lambda\varepsilon\,\cosh\left[\lambda
z_0\right] \qquad
\nonumber \\
& &\  \  +(1-\alpha) \lambda\varepsilon\,\cosh\left[\lambda(1-
z_0)\right].
\label{Eq:distribution-last}
\end{eqnarray}
for $\lambda= \sqrt{s(2+s)}\big/\varepsilon$ and $D_0 =(1+s)\sinh[\lambda]+\lambda \varepsilon \cosh[\lambda]$.
Performing the inverse Laplace transform numerically,
Fig.~\ref{Fig:distribution} shows how this distribution evolves in time. 
For $t \gg {\rm max}[\tau_{\rm mfp},t_{\rm trav}]$
it becomes the gaussian distribution 
$g(N_+,t|z_0) = (2\pi V_t)^{-1/2} \exp\big[-\big(N_+-\langle N_+(t)\rangle\big)^2/(2V_t)\big]$,
where  $\langle N_+(t)\rangle$ and $V_t\equiv{\rm var}[N_+(t)]$
are in Eqs.~(\ref{Eq:result-average}-\ref{Eq:result-covar}).

{\bf Intuitive picture of the results.}
We use the clustering to explain intuitively the 
surprising result that the variances do not decay at large $X/l_{\rm mfp}$.   
We cut a long random-walk into many segments each beginning and ending at $x=X/2$
(see Fig~\ref{Fig:intuitive}a), each taking a time $t_{\rm seg} \sim 3t_{\rm trav}$.
The walk takes a time $t_{\rm trav} \sim X^2/(vl_{\rm mfp})$ to diffuse to a boundary, upon which
it recoils to a distance $l_{\rm mfp}$ from that boundary.
Then the probability that it does {\it not} hit the boundary again before returning to $x=X/2$ is
about $2l_{\rm mfp}/X$.
This probability is {\it tiny}, so the segment contains
$N_{\rm cluster} \sim  X/(2l_{\rm mfp})$ boundary collisions. 
Thus the statistics are similar to 
tossing a coin every time-period $t_{\rm seg}$ 
and saying that a ``head''  is 
$N_{\rm cluster}$ collisions at the L boundary, and a ``tail'' 
is $N_{\rm cluster}$ collisions at the R boundary.
Then $\langle N_{\rm L} \rangle$ and $\langle N_{\rm R}\rangle$ 
go like $N_{\rm cluster} \times t/t_{\rm trav} \sim vt/X$.
However, the variances have $N_{\rm cluster}^2$ in place of $N_{\rm cluster}$, 
and so go like $t /\tau_{\rm mfp}$.
One also has ${\rm covar}[ N_{\rm L},N_{\rm R}] \sim - {\rm var} [ N_{\rm L}] $.
This simple argument gives Eqs.~(\ref{Eq:result-average}-\ref{Eq:result-covar}),
except the ${\cal O}[1]$-prefactors.

{\bf Difference from Smoluchowski.}
Eq.~(\ref{Eq:VMR}) is very different from
Smoluchowski's model of Brownian motion \cite{BrownMotion-book}. 
When estimating the 
number of collisions each liquid particle makes with a macroscopic object 
(the particle undergoing Brownian motion), he neglected correlations, taking 
${\rm var}[N_{\rm L}\! -\!N_{\rm R}] \simeq {\rm var}[N_{\rm L}\! +\!N_{\rm R}] 
\simeq \langle N_{\rm L}\! +\!N_{\rm R}\rangle$. 
Yet, if each liquid particle performs a random walk, 
Eq.~(\ref{Eq:VMR}) shows that is very far from the truth; 
since the container size, $X$, is many orders of magnitude larger than $l_{\rm mfp}$.
In fact, Smoluchowski's assumption  only works  for Brownian motion due to
many-body effects (see below).

{\bf Spin-polarized gases.}  
Such gases, particularly {\He}, are  used as a spin-filter for neutrons \cite{Workshop}, 
a precision magnetometer \cite{Gulob1994,Romalis} 
or to fundamental spin-dependent interactions \cite{Petukhov-PRL-2010,Dubbers}.
They are used for magnetic resonance imaging in medicine \cite{Lung-MRI} 
and engineering \cite{MRI-in-non-medical}.
{\He} gas is typically stored at room temperature and at pressures 0.1-1bar,
in a glass container centimeters across \cite{Walker-Happer-review}; unfortunately, it slowly depolarizes
during storage.
There is a great variety in the quality of the containers;
the gas remains polarized for a few hundred hours in the best containers, 
while  it depolarizes in only a few hours in other superficially identical ones.
The depolarization-processes due to $^3$He-$^3$He scattering in the gas 
\cite{Gamblin1965,Walker-Happer-review,Newbury}
or inhomogeneous external magnetic field  
\cite{Colegrove1963,Schearer-Walter1965,Cates,McGregor,Stoller,Petukhov-PRL-2010,Golub2011,Petukhov-Golub2011,Walker-Happer-review,Clayton2011,Petukhov-Golub2012} differ little between containers.
Thus it is likely that magnetic impurities on the container walls \cite{Bouchiat1960,Walker-Happer-review,Masnou1967,Fitzsimmons1969,Timsit1971,Wu1988,Jacob2001,Jacob2003-04}
are the origin of the huge differences in depolarization times.

\begin{figure}[t]
\includegraphics[width=\columnwidth]{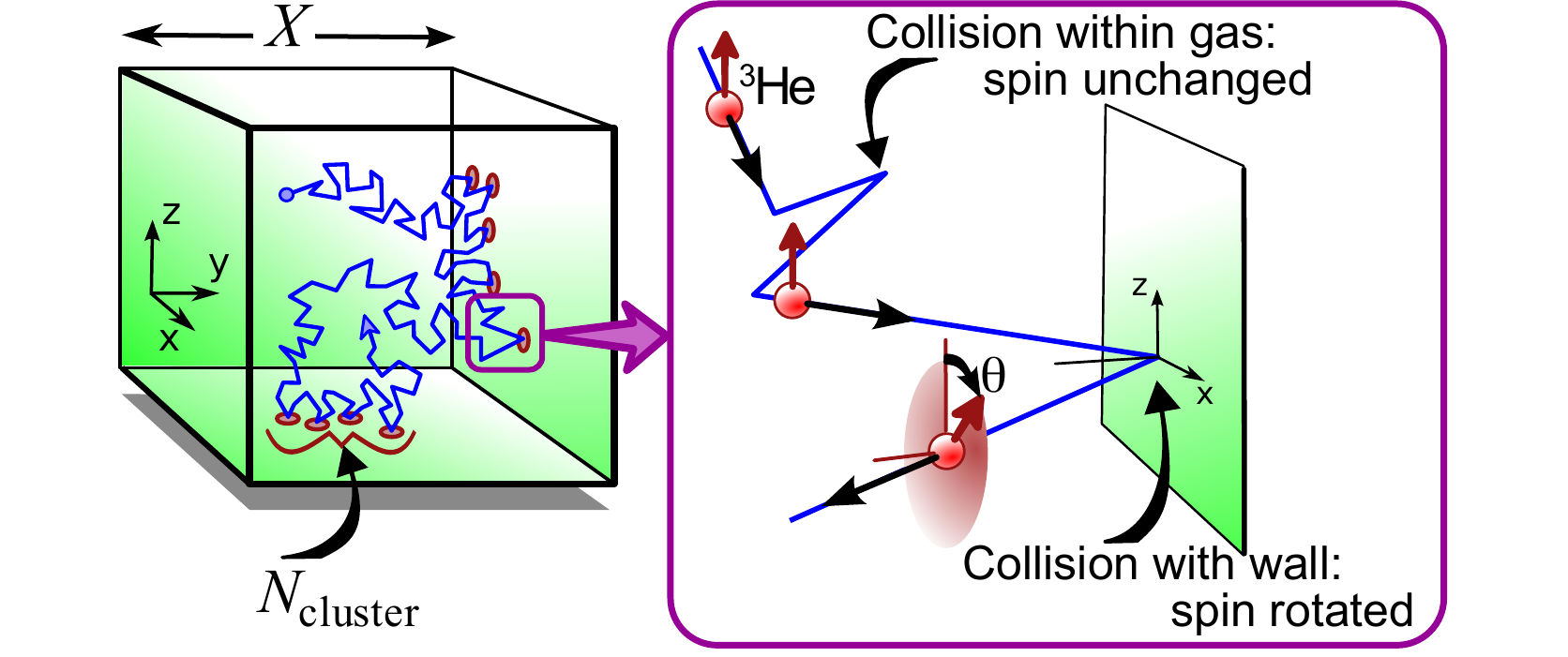}
\caption{\label{Fig:scattering}
The random walk of a {\He} atom 
due to collisions with other {\He} atoms.  
On the left, collisions with the container walls are marked
by ellipses. On the right, we show the 
effect of one such boundary-scattering on the atom's spin.
}
\end{figure}

Here we assume that such magnetic impurities (act on a shorter range than mean free path) slightly rotating the atom's spin each time the atom collides with the walls, see Fig.~\ref{Fig:scattering}.   
Refs.~\cite{Schmiedeskamp2006-part1-2,Schmiedeskamp2006-part3} 
give a microscopic justification of this.
For a typical container (see above) the atomic motion is diffusive with 
$\tau_{\rm mfp}\sim 10^{-10}$s, $X/v\sim 10^{-5}$s, and $t_{\rm trav}\sim 1$s.
The spin-dynamics for $t$ of order the depolarization time (typically tens or hundreds of hours), is deep in the long time regime $t \gg t_{\rm trav} \gg \tau_{\rm mfp}$.
The atom's 3D diffusive motion is very well approximated by three 
uncorrelated 1D random-walks in the $x$, $y$ and $z$-directions.
We take the spin to be rotated by a random angle $\theta_{{\rm L}i}$ at the $i$th collision with the boundary L,
and a random angle $\theta_{{\rm R}i}$ at the $i$ collision with boundary R. 
These angles are tiny, since each atom only depolarizes after very many boundary collisions.
We assume that all rotation are about the same axis 
(relaxing this assumption does not qualitatively change the results \cite{footnote:non-coplanar}).
In this case, the spin-polarization at time $t$ is 
\begin{eqnarray}
S(t)\ \equiv \ \Big<  \,S_0\cos[\Theta(t)]\,\Big> \ =\  S_0 \exp \left[-\half {\rm var}[\Theta(t)] \right],
\label{Eq:polarization}
\end{eqnarray} 
where $\Theta(t) = \sum_{i=1}^{N_{\rm L}(t)} \theta_{Li} + \sum_{i=1}^{N_{\rm R}(t)} \theta_{Ri}$ 
is the total angle that the spin is rotated in a time $t$.
The average in Eq.~(\ref{Eq:polarization}) is over all rotation
angles at each collision and all possible random walks.
Averaging over $\theta_{{\rm L}i}$ and $\theta_{{\rm R}i}$, and then over 
the number of boundary collisions, $N_{\rm L}(t)$ and $N_{\rm R}(t)$, we arrive at
$\langle \Theta(t) \rangle$ and  ${\rm var} [\Theta(t)]$.
For
$\langle \theta \rangle= \langle \theta_{\rm L} \rangle= \langle \theta_{\rm R} \rangle$
and
${\rm var}[\theta] ={\rm var}[\theta_{\rm L}] ={\rm var}[\theta_{\rm R}]$, we have $\langle \Theta(t) \rangle 
= \langle N_+(t) \rangle  \, \langle \theta\rangle$
and 
${\rm var} [\Theta(t)] 
=  \langle N_+(t)\rangle \, {\rm var}[\theta] 
+ \,{\rm var}[N_+(t)] \,\langle \theta\rangle^2$
where $N_+(t)=N_{\rm L}(t)+N_{\rm R}(t)$.
Using Eqs.~(\ref{Eq:result-average}-\ref{Eq:result-covar}), we find the polarization 
decays exponentially at a rate
\begin{eqnarray}
{1 \over T_1}
={ v\, {\rm var}[\theta] \over 2X}
\, +\, { \langle \theta \rangle^2 \over 6\tau_{\rm mfp}},
\label{Eq:inverseT1}
\end{eqnarray} 
for weak enough decay that $T_1 \gg t_{\rm trav}$  \cite{Footnote:ballistic}.
The first term is a typical boundary effect $\propto 1/X$ 
for container size, $X$
\cite{Masnou1967,Fitzsimmons1969,Timsit1971,Wu1988,Jacob2001,Jacob2003-04,Schmiedeskamp2006-part1-2,Schmiedeskamp2006-part3}.
The second term looks like a bulk effect, but is in fact an  {\it $X$-independent boundary effect},
originating from the $X$-independence of Eqs.~(\ref{Eq:result-var},\ref{Eq:result-covar}). 
For typical $^3$He cells $v\tau_{\rm mfp}/X\lesssim 10^{-5}$, so $T_1^{-1}$ is vastly more sensitive 
to the average spin-rotation at each collision, $\langle \theta \rangle$, than the spread 
of the rotations, $\sqrt{{\rm var}[\theta]}$.

Ref.~\cite{Wu1988} had a similar result to Eq.~(\ref{Eq:inverseT1}) for quadrupole fields,
using diffusion equations. 
Ref.~\cite{Petukhov-PRL-2010} got a similar result for magnetic-fields in a region within $\Lambda\ll X$ of the wall, using diffusion and Redfield approximations .  Their result was for $\Lambda \gg l_{\rm mfp}$ (for shorter distances the motion is ballistic and does not obey a diffusion equation). They were surprised to find that Monte Carlo simulations of random walks showed the same behaviour for 
$\Lambda < l_{\rm mfp}$ as for $\Lambda \gg l_{\rm mfp}$.
The origin of these paradoxical boundary-effect 
(which did not decay with increasing system-size), 
was not clear for $\Lambda < l_{\rm mfp}$.  Our above analysis
shows rigorously that such size-independent boundary-effects are a  generic 
property of random walks.

{\bf Comparison with experiments.}
Experiments \cite{Jacob2001,Schmiedeskamp2006-part3} showed
a strong reduction of $T_1$ when the container had previously been placed in a strong magnetic field. 
``Degaussing'' \cite{Jacob2001} the container returned $T_1$ to its original value.
This indicates a low density of magnetic impurities on the walls of their container, 
which the strong field aligned and degaussing un-aligned.
This strong dependence of the depolarization on the history of the cell rules out bulk effects
as the dominant source of depolarization. However, it fits with our model of boundary effects.
 The alignment of different impurities causes $\langle\theta\rangle$ to grow
($\langle\theta\rangle=0$ if they are randomly-oriented), 
and Eq.~(\ref{Eq:inverseT1}) show its extreme sensitivity to the value of this $\langle\theta\rangle$. 
In contrast, the theory in Ref.~\cite{Jacob2001} assumed no correlations 
between an atom's scatterings from the impurities.  So it does not explain
why $T_1$ should depends so strongly on the alignment of the field of different impurities.

Experiments to-date included a uniform magnetic field 
inducing Larmor spin-precession at frequency $\omega \gg t_{\rm trav}^{-1}$. 
Our above results are for the {\it motional narrowing} \cite{Slichter63} regime,  $\om t_{\rm trav} \ll 1$;
and so only give a qualitative explanation of the experiments.
We have, however, applied our method to $\omega  t_{\rm trav}\gg 1$,
and find that $1/T_1$ goes like the square-root of pressure.
This coincides with the prediction for short-range forces  near the walls \cite{Petukhov-PRL-2010}, 
and fits reasonably well to experiments
in Ref.~\cite{Schmiedeskamp2006-part3}.

{\bf Applicability of Smoluchowski's assumption.}
Smoluchowski \cite{BrownMotion-book} applied his assumption (discussed above) to the {\it momentum-transfer} between atoms doing random walks and a macroscopic object (which undergoes Brownian motion as a result).
The depolarization that we discuss can be considered as a {\it spin-transfer} between atoms and a
macroscopic object (the container walls).
Despite superficial similarities, these two are very different.
Inter-atomic scattering in a gas rapidly re-distributes the momentum of any given atom to other atoms, 
making momentum-transfer a  many-body problem.  
Green-Kirkwood-Kubo mean field theory shows that many-body effects typically 
suppress correlations in the momentum-transfer on a timescale of order a few $\tau_{\rm mfp}$.
Thus Smoluchowski's assumption to neglect correlations is not unreasonable.

In contrast, the rate for re-distribution of spin from any given atom to the other atoms
(due to inter-atomic scattering) is tiny; 
 if it were the only depolarization process, then $T_1 > 800$ hours at
gas pressures $\sim$ 1bar. Then boundary collision correlations occur on scales up to 
$t_{\rm trav}\sim 1$s, 
and so are unaffected by many-body effects.  
Thus these correlation are crucial for the spin-transfer 
to the walls, and thus for the depolarization rate.

{\bf Conclusions.} 
For a bounded random walk, we gave the distribution of the number of boundary collisions 
in an arbitrary time. 
Surprisingly, the long time limit exhibits boundary effects which do not decay with increasing system size.  
This could explain the {\He} depolarization rate's extreme sensitivity to
details of the physics at boundaries.
It would be interesting to consider cases where the walk itself has strong correlations, 
e.g.\ sub- or super-diffusive dynamics.

{\bf Acknowledgments.} We are extremely grateful to R.~Golub for important contributions to this work,
and thank G.~Berkolaiko, P.-J.~Nacher and P.~Nozi\`eres for fruitful discussions.



      \setcounter{figure}{0}
      \renewcommand{\thefigure}{S\arabic{figure}}

      \setcounter{equation}{10}

\newpage

\centerline{{\bf SUPPLEMENTARY MATERIAL}}
\vskip 3mm

\begin{figure}[b]
\includegraphics[width=0.43\textwidth]{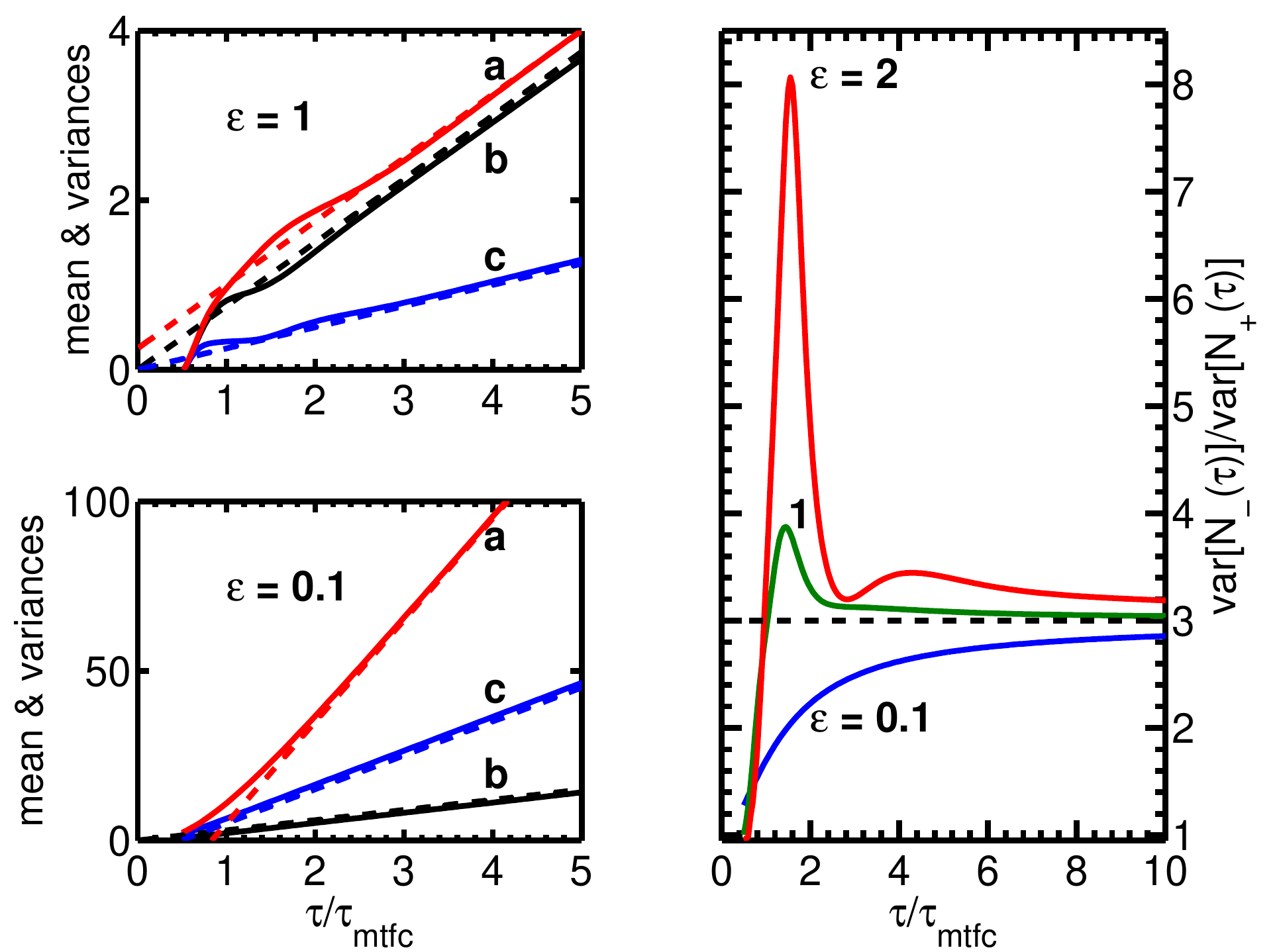}
\caption{\label{Fig:var}
On the left we show the short-time behaviour of ${\rm
var}[N_-(\tau)]$ (quoted "a"), $\langle N_+(\tau)\rangle$ ("b") and
${\rm var}[N_+(\tau)]$ ("c") where
$N_\pm=N_{\rm R}\pm N_{\rm L}$.
We take $a=z_0=1/2$, for which the mean time to first collision (mtfp)
with the boundary,
$\tau_{\rm mtfc}= (2 \varepsilon)^{-1}+(2 \varepsilon)^{-2}$
(so $\tau_{\rm mtfc}=30,\ 0.75,\ 0.11$ for $\varepsilon=0.1,\ 1,\ 5$ repsectively).
On the right we show the variance ratio for various $\varepsilon$. 
Dashed  lines are the long-time asymptotes given by Eqs.~(\refEqresultaverage-\refEqresultcovar).
On the left we improve these asymptotes by adding small $\tau$-independent terms ${\cal O}[1,\varepsilon^{-2}]$ as fit parameters.
}
\end{figure}

{\bf Derivation of $N_{\rm L,R}$ statistics.}
Here we give the calculation we outlined in the ``Model and Results'' section
of the paper, using techniques in Ref.~\citeFellerprobbook.
For a probability density $F(N_{\rm L},N_{\rm R},\tau|z_0)$, 
that the walker hits the L and R boundaries $N_{\rm L}$ and $N_{\rm R}$ times in time $\tau$, we
have
$\Psi(\tau|z_0) = \sum_{N_{\rm L},N_{\rm R}=0}^{\infty}\mu_{\rm L}^{N_{\rm L}}
\mu_{\rm R}^{N_{\rm R}}\,F(N_{\rm L},N_{\rm R},\tau|z_0)$.
Then 
\begin{eqnarray}
\left[ {\rmd \Psi /\rmd \mu_W} \right]_{\mu_{\rm L,R}=1} &=& \langle N_W\rangle,
\nonumber \\
\left[ {\rmd^2 \Psi /\rmd \mu_W^2} \right]_{\mu_{\rm L,R}=1} &=&
\langle N_W(N_W-1)\rangle,
\nonumber \\
\left[ {\rmd^2 \Psi  / \rmd \mu_{\rm L} \rmd \mu_{\rm R}}  \right]_{\mu_{\rm
L,R}=1} &=& \langle N_{\rm L} N_{\rm R}\rangle,
\label{Eq:moments-as-derivatives}
\end{eqnarray}
for $W \in \{ {\rm L},{\rm R}\}$.
To find $ \Psi$ and its derivatives,
we work in Laplace space, 
defining the Laplace transform of any function $f(\tau)$ as 
$\widehat{f}(s)=\int_{0}^{\infty}f(\tau)\e^{-s\tau} {\rm d}\tau$.  
We multiply Eq.~(\refEqtelegrapher)
by $\e^{-st}$ and integrate over $t$ from $0$ to $\infty$, 
noting that $\int_{0}^{\infty} {\rm d}\tau \, \e^{-s\tau} {{\rm d} \over {\rm d} \tau} P_\pm(z,\tau|z_0) = P_\pm(z,\tau=0|z_0) 
+ s \widehat{P}_\pm(z,s|z_0)$. Defining the vector 
\begin{eqnarray}
\widehat{{\bf P}} (z,s|z_0) =\left( \begin{array}{c} 
\widehat{P}_+ (z,s|z_0) \\ 
\widehat{P}_- (z,s|z_0) 
\end{array}\right),
\end{eqnarray}
we have 
\begin{eqnarray}\label{tel2}
\frac{d}{dz} \widehat{{\bf P}} (z,s|z_0)
&=&{\mathbb M}(s)  \, \widehat{{\bf P}} (z,s|z_0) +{\bf F}\,\delta(z-z_0), \qquad
\label{Eq:P(s)-diff-eqn}
\end{eqnarray}
where 
\begin{eqnarray}
{\mathbb M}(s)=  {1\over \varepsilon} \left( \! \begin{array}{cc} -1-s & 1\\ -1 & 1+s\end{array} \! \right), \quad
{\bf F}= {1\over \varepsilon} \left( \!\! \begin{array}{c} a \\ a-1 \end{array} \!\! \right).\quad
\end{eqnarray}
The boundary conditions are
${\bf b}^{\rm T}_{\rm L}\,\widehat{{\bf P}}(0,s |z_0) =0$ and $ {\bf b}^{\rm T}_{\rm R}\,\widehat{{\bf P}}(1,s |z_0) = 0$, 
where T indicates a transpose of
\begin{eqnarray}
{\bf b}_{\rm L} = \left( \! \begin{array}{c} 1 \\ -\mu_{\rm L}\end{array}\! \right),
\quad
{\bf b}_{\rm R} = \left( \! \begin{array}{c} \mu_{\rm R} \\ -1\end{array}\! \right).
\label{Eq:vector-bondary-cond}
\end{eqnarray}

The Laplace transform of the survival probability is
$\widehat{\Psi}(s|z_0) = \int_0^1 {\rm d}z \big( \widehat{P}_+(z,s|z_0)  + \widehat{P}_-(z,s|z_0) \big)$.
Subtracting the first line of the matrix equation in Eq.~(\ref{tel2}) from the second line,
gives
$\widehat{P}_+ + \widehat{P}_- = -s^{-1} \big( \varepsilon {{\rm d} \over {\rm d}z} [ \widehat{P}_+ -\widehat{P}_-] -\de (z-z_0) \big) $.  
Placing this in the integral makes it easy to evaluate, 
then using the boundary conditions we get
\begin{eqnarray}
\widehat{\Psi}(s|z_0) = s^{-1} \big[1
- (1-\mu_{\rm L})\varepsilon \widehat{P}_{\rm L-}
- (1-\mu_{\rm R})\varepsilon \widehat{P}_{\rm R+} \big], \quad
\label{Eq:Psi(s)}
\end{eqnarray}
where we use the shorthand 
$\widehat{P}_{\rm L-}\equiv  \widehat{P}_{-}(z=0,s|z_0)$ and 
$\widehat{P}_{\rm R+}\equiv  \widehat{P}_{+}(z=1,s|z_0)$.

Thus, to find $\widehat{\Psi}(s|z_0)$, we do not need $\widehat{{\bf P}}(z,s|z_0)$ for all $z$, 
we only need  its values at the boundaries $z=0,1$.
In what follows, we refer to these values using the shorthand $\widehat{{\bf P}}_{\rm L}\equiv \widehat{{\bf P}}(z=0,s|z_0)$ 
and $\widehat{{\bf P}}_{\rm R}\equiv \widehat{{\bf P}}(z=1,s|z_0)$.
 To get these, we define 
the Fourier transform of $\widehat{f}(z,s)$ as $\widetilde{f}(k,s) = \int_0^1 {\rm d } z \, \e^{\rmi kz} \widehat{f}(z,s)$. 
We apply this to Eq.~(\ref{Eq:P(s)-diff-eqn}), using
$ \int_0^1 {\rm d } z \, \e^{\rmi kz} \frac{d}{dz} \widehat{{\bf P}} (z,s|z_0) =
  \e^{\rmi k} \widehat{{\bf P}}_{\rm R} -  \widehat{{\bf P}}_{\rm L} -\rmi k  \widetilde{{\bf P}}(k,s|z_0) $.
This gives an equation for  $\widetilde{{\bf P}}(k,s|z_0)$,
which we write in the basis where ${\mathbb M}$ is diagonal,
using
\begin{eqnarray}
\ {\mathbb M}={\mathbb V}^{-1} \left( \! \begin{array}{cc} \lambda  & 0 \\ 0 & -\lambda \end{array} \! \right) {\mathbb V},
\quad{\mathbb V}\! = \left( \! \begin{array}{cc} c_+(1+s -\varepsilon \lambda)\  & -c_+ \\
c_-(1+s +\varepsilon \lambda)\  & -c_-  \end{array} \! \right),
\nonumber 
\end{eqnarray}
where $\pm\lambda$ are  ${\mathbb M}$'s eigenvalues, with  $\varepsilon\lambda=\sqrt{s(2+s)}$, and 
$c_\pm= [(1+s \mp \varepsilon\lambda)^2-1]^{-1/2}$.
The result is
\begin{eqnarray}
0  &=& \left(\! \begin{array}{cc} \lambda +\rmi k & 0\\ 0 & -\lambda +\rmi k\end{array} \!\right)
{\mathbb V}\,\widetilde{{\bf P}}(k,s|z_0) 
\nonumber \\ 
& & \qquad \qquad \ +\,\e^{\rmi kz_0}\,{\mathbb V}\,{\bf F} +\,{\mathbb V}\,\widehat{{\bf P}}_{\rm L}
- \,\e^{\rmi k} \,{\mathbb V}\,\widehat{{\bf P}}_{\rm R}  , \qquad
\label{Eq:Fourier-transformed-eqn}
\end{eqnarray}
Eq.~(\ref{Eq:Fourier-transformed-eqn}) is true for all $k$, so it must be true for $k=\rmi \lambda$.
Then $\widetilde{{\bf P}}(k,s|z_0)$ drops out of the upper elements in this vector equation
(assuming $\widetilde{{\bf P}}$ does not diverge at $k=\rmi \lambda$), 
leaving an equation for 
$\widehat{{\bf P}}_{\rm L}$ and $\widehat{{\bf P}}_{\rm R}$.
For $k=-\rmi \lambda$, $\widetilde{{\bf P}}(k,s|z_0)$ drops out of  the  lower elements,
giving us a second equation for $\widehat{{\bf P}}_{\rm L}$ and $\widehat{{\bf P}}_{\rm R}$.
These equations are
\begin{eqnarray}
\e^{-\lambda} \ {\bf v}_+^{\rm T} \, \widehat{{\bf P}}_{\rm R} \ -\  {\bf v}_+^{\rm T} \,\widehat{{\bf P}}_{\rm L}
&=&\e^{-\lambda z_0} \,  {\bf v}_+^{\rm T}\,{\bf F},
\label{Eq:Simultaneous1-a}
\\
\e^{\lambda} \  {\bf v}_-^{\rm T} \,\widehat{{\bf P}}_{\rm R} \ -\   {\bf v}_-^{\rm T}\,\widehat{{\bf P}}_{\rm L} 
&=& \e^{\lambda z_0}  {\bf v}_-^{\rm T}\,{\bf F}, 
\label{Eq:Simultaneous1-b}
\end{eqnarray}
where we define the vectors  
${\bf v}_+^{\rm T} \equiv (1,0){\mathbb V} = c_+(1+s-\varepsilon\lambda \   ,\,-1)$ and 
${\bf v}_-^{\rm T} \equiv (0,1){\mathbb V} = c_-(1+s+\varepsilon\lambda \   , \, -1)$.
Eqs.~(\ref{Eq:Simultaneous1-a},\ref{Eq:Simultaneous1-b}) contain four unknowns;
$\widehat{P}_{\rm L+}$, $\widehat{P}_{\rm L-}$, $\widehat{P}_{\rm R+}$, $\widehat{P}_{\rm R-}$.
However the boundary conditions, Eq.~(\ref{Eq:vector-bondary-cond}), enable us to write 
$\widehat{{\bf P}}_{\rm R}  = {\bf u}_{\rm R}\,\widehat{P}_{\rm R+} $ 
and 
$\widehat{{\bf P}}_{\rm R}  = {\bf u}_{\rm L}\,  \widehat{P}_{\rm L-}$, 
where 
\begin{eqnarray}
{\bf u}_{\rm R}  =\left(\! \begin{array}{c} 1 \\ \mu_{\rm R}\end{array} \! \right),
\qquad 
{\bf u}_{\rm L}  = \left(\! \begin{array}{c} \mu_{\rm L} \\ 1\end{array} \! \right),
\end{eqnarray}
and $\widehat{P}_{\rm R+},\widehat{P}_{\rm L-}$ are the two quantities we need for 
Eq.~(\ref{Eq:Psi(s)}).
Substituting this into Eqs.~(\ref{Eq:Simultaneous1-a},\ref{Eq:Simultaneous1-b}),
we get  equations for $\widehat{P}_{\rm R+}$ and $\widehat{P}_{\rm L-}$, which in matrix form
are
\begin{eqnarray}
\left( \! \begin{array}{cc}  
\e^{-\lambda} {\bf v}_+^{\rm T}{\bf u}_{\rm R}   & -{\bf v}_+^{\rm T} {\bf u}_{\rm L}   \\
 \e^{\lambda} {\bf v}_-^{\rm T}{\bf u}_{\rm R}   & -{\bf v}_-^{\rm T} {\bf u}_{\rm L}    
 \end{array} \! \right)\!
\left( \! \begin{array}{c}   
\widehat{P}_{\rm R+} \\  
 \widehat{P}_{\rm L-} 
 \end{array} \! \right) 
 \! &=&\!
  \left( \! \begin{array}{c}  
\e^{-\lambda z_0} {\bf v}_+^{\rm T} {\bf F} \\  
\e^{\lambda z_0} {\bf v}_-^{\rm T} {\bf F}  \end{array} \! \right).\qquad\ \
\end{eqnarray}
Inverting this equation we find that
\begin{eqnarray}
\widehat{P}_{\rm R+} 
\!&=&\! { \e^{\lambda z_0} [{\bf v}_+^{\rm T} {\bf u}_{\rm L}][{\bf v}_-^{\rm T} {\bf F}]
- \e^{-\lambda z_0} [{\bf v}_-^{\rm T} {\bf u}_{\rm L} ] [{\bf v}_+^{\rm T} {\bf F}]
\over
 \e^{\lambda}  [{\bf v}_+^{\rm T} {\bf u}_{\rm L} ][{\bf v}_-^{\rm T} {\bf u}_{\rm R}]
- \e^{-\lambda}  [{\bf v}_-^{\rm T} {\bf u}_{\rm L} ][{\bf v}_+^{\rm T} {\bf u}_{\rm R}]
},
\nonumber \\
\widehat{P}_{\rm L-} 
\!&=& \!{ 
\e^{-\lambda (1-z_0)}  [{\bf v}_+^{\rm T} {\bf u}_{\rm R}] [{\bf v}_-^{\rm T} {\bf F}]
 -\e^{\lambda (1-z_0)} [{\bf v}_-^{\rm T} {\bf u}_{\rm R}][ {\bf v}_+^{\rm T} {\bf F}]
 \over
 \e^{\lambda}   [{\bf v}_+^{\rm T} {\bf u}_{\rm L}][ {\bf v}_-^{\rm T} {\bf u}_{\rm R}]
- \e^{-\lambda}  [{\bf v}_-^{\rm T} {\bf u}_{\rm L}][{\bf v}_+^{\rm T} {\bf u}_{\rm R}]
}.
\nonumber
\end{eqnarray}
Here
${\bf v}_\pm^{\rm T} {\bf u}_{\rm R} = 1+s -\mu_{\rm R} \mp \varepsilon \lambda$, 
${\bf v}_\pm^{\rm T} {\bf u}_{\rm L} = (1+s)\mu_{\rm L}-1 \mp \mu_{\rm L} \varepsilon \lambda$,
and 
${\bf v}_\pm^{\rm T}{\bf F} =  \varepsilon^{-1} (1+as \mp  a \varepsilon\lambda)$,
so we get
 \begin{eqnarray}
\widehat{P}_+(1,s | z_0)
\! &=&\!  {A^+_{\rm L} \sinh [\lambda z_0] +B^+_{\rm L} \lambda \varepsilon 
\cosh [\lambda z_0] 
\over 2\varepsilon D}, 
\nonumber \\
\widehat{P}_-(0,s|z_0) 
\! &=& \! {A^-_{\rm R} \sinh [\lambda -\lambda z_0] 
     + B^-_{\rm R} \lambda \varepsilon \cosh [\lambda -\lambda z_0] 
\over 2\varepsilon D},
\nonumber 
\end{eqnarray}
where 
$A^\pm_W = (1-\mu_W)(s+2) \mp (1-2a)(1+\mu_W)s$,
$B^\pm_W = (1+\mu_W)\mp(1-2a)(1-\mu_W)$
and 
$D=
\left[(1-\mu_{\rm R})(1-\mu_{\rm L}) +s(1+\mu_{\rm L}\mu_{\rm R})\right]\,\sinh[\lambda] 
+ (1-\mu_{\rm L}\mu_{\rm R})\lambda\varepsilon \cosh[\lambda]$.
Substituting these into Eq.~(\ref{Eq:Psi(s)}), 
we get the Laplace transformed survival probability, $\widehat{\Psi}(s|z_0)$.

\begin{figure}[t]
\includegraphics[width=6.4cm]{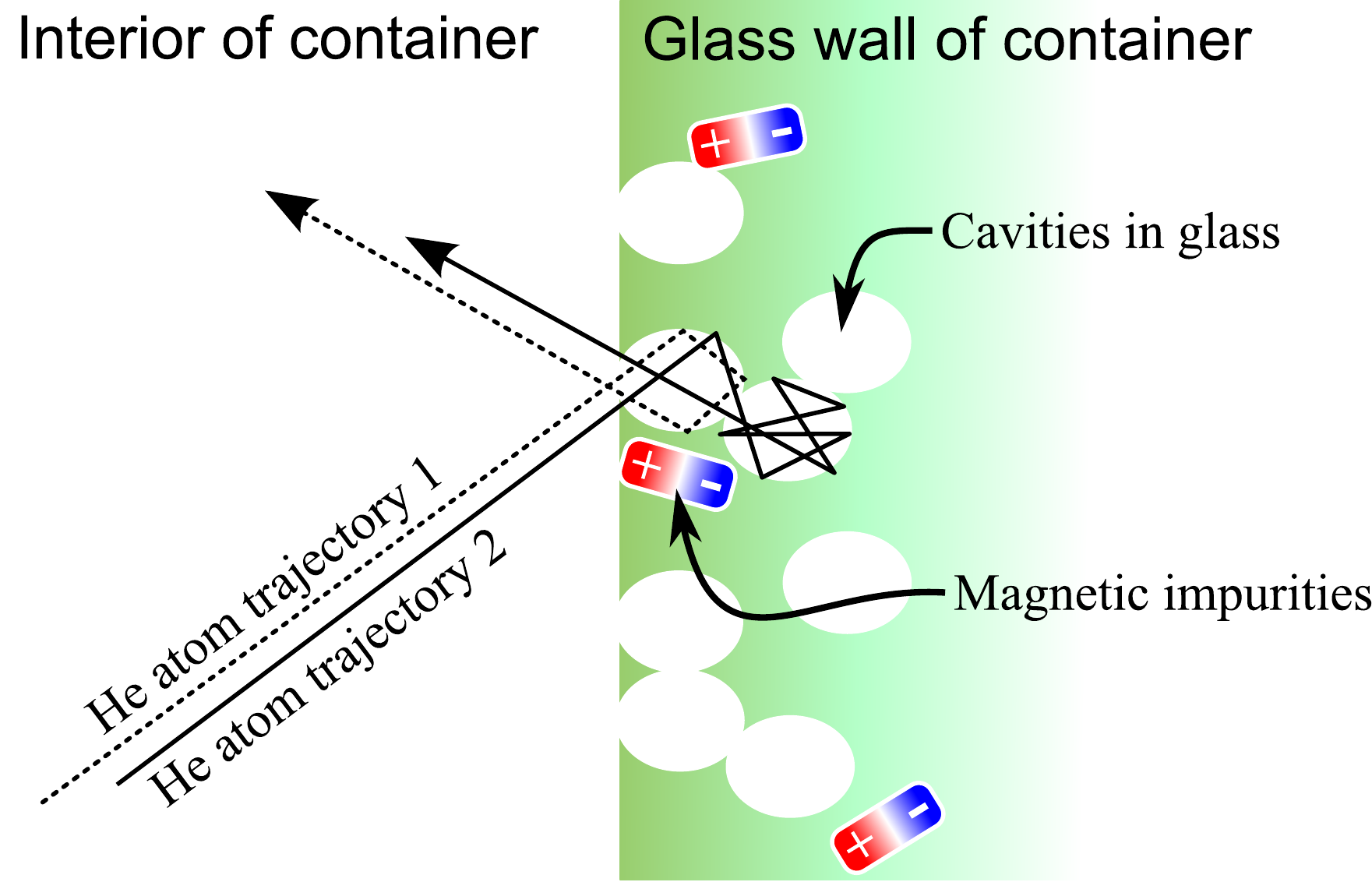}
\caption{\label{Fig:scattering-details} 
A {\He} atom hitting the wall may enter microscopic cavities 
in the wall, getting trapped there for some time.  This trapping may be geometric 
(bouncing as shown) or chemical (van der Waals bonding to the glass). 
The trapping time is random, and uncorrelated between different collisions 
at the same place on the wall (e.g.~longer for trajectory 2 than 1),
The spin's rotation $\theta$ is given by the field
at the trapping site  (due to magnetic impurities), and the trapping time.
}
\end{figure}

Laplace transforming Eqs.~(\ref{Eq:moments-as-derivatives}) we evaluate the derivatives exactly.
It is convenient to define $N_\pm(s|z_0)= N_{\rm R} (s|z_0)\pm N_{\rm L}(s|z_0)$, 
for which we have
\begin{eqnarray}
\big\langle\widehat{N}_{\pm}(s|z_0)\big\rangle
 \! &=& \!
\frac{(1-2a)s\left(\sinh\left[\lambda
(1-z_0)\right]\mp\sinh\left[\lambda z_0\right]\right)
}{2s^2\,\sinh[\lambda]}
\nonumber\\
&+&\!
\frac{\lambda\varepsilon\left(
\cosh\left[\lambda
(1-z_0)\right]\pm\cosh\left[\lambda z_0\right]\right)
}{2s^2\,\sinh[\lambda]}, \quad\quad
\label{Eq:N_pm}
\end{eqnarray}
\begin{eqnarray}
& & \hskip -7mm
\big\langle \widehat{N}_{\pm}^2(s|z_0) \big\rangle 
\nonumber \\
&=&\!
\left(\frac{(1\pm 1)\lambda\varepsilon\cosh\left[\lambda\right]
}{s\,\sinh[\lambda]} \mp 1\right)
\big\langle\widehat{N}_+(s|z_0)\big\rangle
\nonumber\\
&\pm& \!
\frac{\lambda\varepsilon
\left(a\cosh\left[\lambda
(1-z_0)\right]+(1-a)\cosh\left[\lambda z_0\right]\right)
}{s^2\,\sinh[\lambda]}
\nonumber\\
&\mp& \!
 \frac{(1+as)\sinh\left[\lambda
(1-z_0)\right]+(s+1-as)\sinh\left[\lambda z_0\right]
}{s^2\,\sinh[\lambda]},  \, \qquad 
\\
& & \hskip -7mm
\big\langle\widehat{N}_+(s|z_0)\widehat{N}_-(s|z_0)\big\rangle
\nonumber \\
 &=& \!
\big\langle\widehat{N}_-(s|z_0)\big\rangle
+
\frac{s \sinh[\lambda]-\lambda\varepsilon \cosh[\lambda]}{2s^2\sinh^2[\lambda]}
\nonumber \\
& & \qquad\qquad  \times \Big(
(1-2a)s\big( \sinh[\lambda z_0] -\sinh[\lambda(1-z_0)]\big) 
\nonumber \\
& & \qquad\qquad\quad
-\lambda\varepsilon \big(  \cosh[\lambda z_0] +\cosh[\lambda(1-z_0)]  \big) \Big).
\label{Eq:N_+N_-}
\end{eqnarray}
To write these in terms of the number of L and R boundary collisions,
we use
$\big\langle \widehat{N}_{\rm R}(s)\big\rangle = \half \big\langle \widehat{N}_+(s)\big\rangle +\half\big\langle \widehat{N}_-(s)\big\rangle$,
$\big\langle \widehat{N}_{\rm L}(s)\big\rangle = \half \big\langle \widehat{N}_+(s)\big\rangle -\half\big\langle \widehat{N}_-(s)\big\rangle$,
while
\begin{eqnarray}
& &\hskip -4mm \big\langle \widehat{N}_{\rm R}^2(s)\big\rangle 
= \quarter\big\langle \widehat{N}_+^2(s)\big\rangle + \quarter\big\langle \widehat{N}_-^2(s)\big\rangle
+\half\big\langle \widehat{N}_+(s) \widehat{N}_-(s)\big\rangle, \qquad
\\
& &\hskip -4mm  \big\langle \widehat{N}_{\rm L}^2(s)\big\rangle 
= \quarter\big\langle \widehat{N}_+^2(s)\big\rangle +\quarter\big\langle \widehat{N}_-^2(s)\big\rangle
-\half\big\langle \widehat{N}_+(s) \widehat{N}_-(s)\big\rangle,
\\
& &\hskip -4mm  \big\langle \widehat{N}_{\rm L}(s)\widehat{N}_{\rm R}(s) \big\rangle 
= \quarter\big\langle \widehat{N}_+^2(s)\big\rangle -\quarter\big\langle \widehat{N}_-^2(s)\big\rangle.
\end{eqnarray}
where for compactness we do not show explicitly that all these 
quantities depend on $z_0$.

For $s\ll {\rm min}[1,\varepsilon^2]$, one can perform a small $s$ expansion (we must keep terms up to $s^{-2}$)
and  perform the inverse Laplace transform, getting the results for $\tau \gg {\rm max}[1, \tau_{\rm trav}]$.  We give these results in
Eqs.~(\refEqresultaverage-\refEqresultcovar), neglecting $t$-independent terms of  ${\cal O}[1]$ and ${\cal O}[\varepsilon^{-2}]$,
equivalent to a small shift of time, $t \to t+{\cal O}[\tau_{\rm mfp},t_{\rm trav}]$.
Fig.~\ref{Fig:var} shows the finite-time behavior of these quantities,
found by numerically inverting the Laplace transform of  Eq.~(\ref{Eq:N_pm}-\ref{Eq:N_+N_-}).
They  go  fairly rapidly to their long-time linear-$\tau$ behavior, with 
${\rm var}[N_-(\tau|z_0)]/{\rm var}[N_+(\tau|z_0)]\to 3$ for long-times.

The mean time to first collision (mtfc) is 
$\tau_{\rm mtfc}=\int_0^\infty \rmd \tau \, \tau \, \big( {\rm d} F(0,0,\tau|z_0)/{\rm d}\tau\big) =\widehat{\Psi}(0|z_0)|_{\mu_{\rm L,R}=0}$, and has a simple form;
$\tau_{\rm mtfc}=\tau_{\rm ball}+\tau_{\rm diff}$,
where the times in the ballistic and diffusive limits are $\tau_{\rm ball}= [a(1-z_0)+(1-a)z_0]/\varepsilon$
and $\tau_{\rm diff}= z_0(1-z_0)/(\varepsilon^2)$.
The typical time to traverse the system, $\tau_{\rm trav} \simeq 2\tau_{\rm mtfc}(z_0=1/2)$;
which is $t_{\rm trav}\simeq X/v + X^2/ (v^2 \tau_{\rm mfp})$ in dimensionful units.

We conclude by pointing out  that we never needed the solution of Eq.~(\ref{Eq:P(s)-diff-eqn}).
However for completeness we point out that it is
$\widehat{{\bf P}}(z,s|z_0) 
= \e^{z{\mathbb M}(s)} \Big[  \widehat{{\bf P}}(z=0,s|z_0) + \e^{-z_0{\mathbb M}(s)}{\bf F}\, \theta(z\!-\!z_0) \Big] $,
where $\theta(z)$ is a Heaviside step function, and the
vector $\widehat{{\bf P}}(z=0,s|z_0)$ is given above.

{\bf Intermediate times in diffusive regime.}
In the diffusive regime $v\tau_{\rm mfp} \ll X$, we can look at the  intermediate $t$ regime, defined by
$\tau_{\rm mfp} \ll t \ll \tau_{\rm trav}$.   
We follow the above derivation, but now consider
$\varepsilon^2 \ll s \ll 1$. Then the collision statistics are very  
sensitive to the walker's  initial position, $x_0$; 
if $x_0$ is too far from a boundary, there are no collisions.
Averaging uniformly over $x_0$, we find 
$\big\langle N_{\rm L}\big\rangle$
and $\big\langle N_{\rm R}\big\rangle$ are as in Eq.~(\refEqresultaverage), while
\begin{eqnarray}
{\rm var}[N_{\rm L}] \ = \ 
{\rm var}[N_{\rm R}] &=& \, v\, t^{3/2}/(X\,\tau_{\rm mfp}^{1/2}), \qquad
\label{Eq:result-var-intermediate-time}
\\
{\rm covar}[N_{\rm L},N_{\rm R}] \,\simeq  - \,\big\langle N_{\rm L}\big\rangle^2 
&=&  -(vt/X)^2 .
\label{Eq:result-covar-intermediate-time}
\end{eqnarray}
Thus they decay with increasing system-size,
with the covariance having a much smaller magnitude than the variances.
Unlike in the long time limit, the distribution is not gaussian
see Fig.~{\refFigdistribution}, and the 
typical fluctuations, $\sqrt{{\rm var}[N_{\rm L,R}] }$, are much larger than the averages, 
$\langle N_{\rm L,R}\rangle$. 
As a result the variance gives limited information about the nature of the distribution.

\begin{figure}[t]
\includegraphics[width=5.6cm]{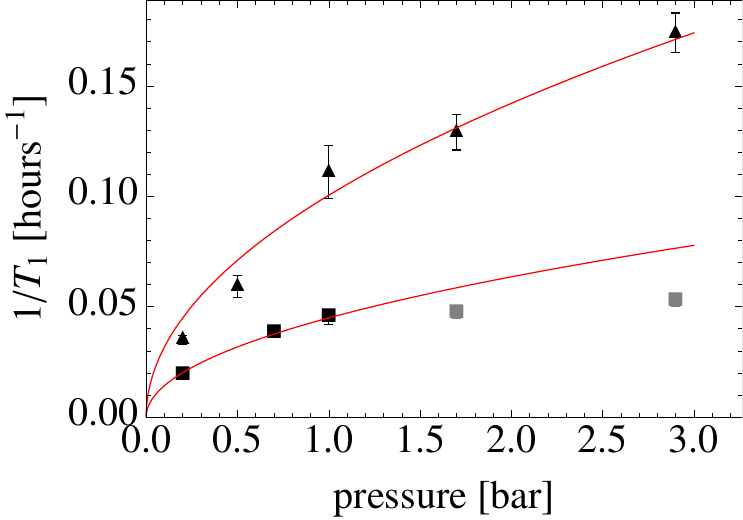}
\caption{\label{Fig:expt} 
Data from Fig.~6 of Ref.~{\citeSchmiedeskamppartthree} (squares are sample C\#4 
and triangles are C\#13),
with $\alpha p^{1/2}$ (our theory),
where $\alpha$ is chosen for the best fit.
We do not include the expected crossover to a $1/p$ behavior at higher $p$;
instead we do not fit to the points where
this effect may be significant  (two points marked in gray for sample C\#4).   
The remaining points are as well fitted by our $\sqrt{p}$ prediction as by 
the fitting in  Ref.~{\citeSchmiedeskamppartthree}
(a theory introduced in Ref.~{\citeJacob} which is linear in $p$ at small $p$,
with the crossover to $1/p$ at higher $p$).
More experiments are necessary to distinguish between the theories.
}
\end{figure}

{\bf Justification of our model for {\He}.}
In experiments {\citeJacobSchmiedeskamppartthree}, 
boundary-induce depolarization is likely due to 
localized magnetic impurities, see Fig.~\ref{Fig:scattering-details}.
However at each collision with a given region of the 
boundary, the atom will be trapped there {\citeSchmiedeskamppartonetwo} for a different time 
(see  Fig.~\ref{Fig:scattering-details}), thereby acquiring a different rotation angle.
As a result, the rotation angle is not directly related to the position at which the atom hits the boundary.
Thus it seems a reasonable simplification to assume all scattering at a given
boundary by a single distribution of angles averaged over that surface. 
We make this simplification in our model,
characterizing the distribution by $\langle \theta \rangle$ and ${\rm var}[\theta]$.
The central message of our discussion of polarized-gases was that the depolarization 
rate is crucially dependent on the correlations between boundary scatterings.
At worst, the above argument under-estimates such correlation effects, by neglecting 
correlation due by multiple scatterings in the same region of the boundary.

{\bf Depolarization with Larmor precession.}
The spin of {\He} in an external magnetic field precesses at a rate $\omega$.
This was about  $10^5{\rm s}^{-1}$ for the experiments in 
Refs.~{\citeJacobSchmiedeskamppartthree}
(the other parameters  were
$\tau_{\rm mfp}\sim 10^{-10}$s, $X/v\sim 10^{-5}$s, and $t_{\rm trav}\sim 1$s). 
Thus we want the depolarization rate, $T^{-1}_1$ for 
$ \tau_{\rm mfp} \ll 1/\omega \ll t_{\rm trav}$.
Here $T_1^{-1}$ is the rate of relaxation along the axis of the external field 
(the perpendicular decay rate, $T^{-1}_2$, 
is much faster).   

We argue that  $\omega^{-1}$ acts as a cut-off on the correlations of the rotation angle.
If the spin rotates clockwise about the $y$-axis at every boundary collision, the angle to the 
$z$-axis (the external field axis) will systematically grow for times $\ll \omega^{-1}$,
however after a time of order $(2 \omega)^{-1}$ the spin will have precessed $180^\circ$ about the
$z$-axis.  At this point clockwise rotations about the $y$-axis will {\it reduce} the angle between the spin 
and the $z$-axis.   With this cut-off at  $\omega^{-1}$,
 the variance in the rotation angle is
the sum of the variances  acquired in each time-slice of order $\omega^{-1}$, so
${\rm var}[\Theta(t)] \sim \omega t \big[ \big\langle N_+(\omega^{-1})\big\rangle\, {\rm var}[\theta ]  
+  {\rm var}[N_+(\omega^{-1})]  \,\langle \theta\rangle^2 \big]$.
We substitute in 
Eqs.~(\ref{Eq:result-var-intermediate-time},\ref{Eq:result-covar-intermediate-time}), 
and get the depolarization rate 
for $\tau_{\rm mfp} \ll \omega^{-1} \ll t_{\rm trav}$, 
\begin{eqnarray}
T_1^{-1} \sim  {v \over X} \left[ {\rm var}[\theta ]
+ (\omega \tau_{\rm mfp})^{-1/2} \big\langle \theta \big\rangle^2  \right].
\label{Eq:inverseT1-finite-fields}
\end{eqnarray}
Hence in this regime, $1/T_1$ decays with increasing system-size, $X$.
Since $\tau_{\rm mfp}^{-1}\propto$ gas pressure, $p$, 
the second term goes like $\sqrt{p}$.
There is no experimental consensus on the $p$-dependence in such systems 
{\citeJacobSchmiedeskamppartthree}.
While we do not know all the experimental conditions, 
the data in Fig.~6 of Ref.~{\citeSchmiedeskamppartthree}
can be fitted with a $\sqrt{p}$, see Fig.~\ref{Fig:expt}.

For typical experimental parameter, the prefactor on the 
second term in Eq.~(\ref{Eq:inverseT1-finite-fields})
is of order $10^{5/2} \sim 300$ times larger than that of the first term.
Compared with the case with no external magnetic field (where the prefactor on the second term was 
$10^5$ times larger than that of the first term), the magnitude of the second term is greatly reduced.  This is a consequence of the fact that correlations are absent on timescales larger than $\om^{-1}$. 
None the less, the remaining correlations are sufficient to make $1/T_1$ {\it much} more sensitive to 
average angle, $\langle \theta\rangle$, than to the 
spread of angles, $\sqrt{{\rm var}[\theta]}$.



\begin{thebibliography}{1}
\bibitem{footnote:non-coplanar}
Alternatively, one could take the rotation to be about a randomly chosen axis.
Then the average rotation at each collision will only be non-zero (analogous to
$\langle \theta_{\rm L,R}\rangle \neq 0$ in the calculation in this article),
if the distribution for this randomly chosen axis is non-uniform.
\end{thebibliography}
\end{document}